\documentclass[]{iopart}

 \usepackage[english]{babel}
 \usepackage{color}
 \usepackage{pstricks}
 \usepackage{bm}
\usepackage{iopams}

\usepackage{bbold}
\usepackage{graphicx}
\usepackage{cite}

\begin{document}

\title[Three interacting atoms in a one-dimensional 
trap]{Three interacting atoms in a one-dimensional trap:\\
A benchmark system for computational approaches}

\author{Pino D'Amico and Massimo Rontani}
\eads{\mailto{pino.damico@nano.cnr.it}, \mailto{massimo.rontani@nano.cnr.it}}
\address{CNR-NANO Research Center S3, Via Campi 213/a, 41125 Modena, Italy}



\begin{abstract}
We provide an accurate calculation of the energy spectrum
of three atoms interacting through a contact force
in a one-dimensional harmonic trap, considering both spinful
fermions and spinless bosons.
We use fermionic energies as a benchmark for 
exact-diagonalization technique (also known as full
configuration interaction), which is found to slowly converge
in the case of strong interatomic attraction. 
\end{abstract}

\submitto{\JPB}

\maketitle

\section{Introduction}

Experimental advances allow us to confine a chosen number of few quantum
degenerate atoms in a trap with unit precision.
Whereas this capability was first demonstrated for bosons
in optical lattices \cite{Cheinet2008,Will2010},
the Heidelberg group
applied a spilling technique to subtract $^6$Li fermionic atoms one by one
from a single trap,
down to the zero limit \cite{Serwane2011,Zuern2012,Zuern2013,Wenz2013}.
Due to
the strong anisotropy of the magneto-optical trap employed,
the few-atom system was effectively one-dimensional.
The tunability of both confinement potential and interaction---the
latter achieved by sweeping a magnetic offset field through a Feshbach
resonance \cite{Chin2010}---opens new avenues
to understand interacting particles in one dimension
\cite{Giamarchi2003,Giorgini2008,Bloch2008,Deshpande2010,Ilani2010,Cazalilla2011,Guan2013}.

So far the experiments \cite{Serwane2011,Zuern2012,Zuern2013,Wenz2013}
have revealed that phenomena
previously investigated in the presence of many atoms may be studied 
in the limit of few particles as well,
as the Fermi-Bose duality
\cite{Kinoshita2004,Paredes2004,Zuern2012}
(also known as `fermionization' \cite{Girardeau1960,Cheon1999}),
the formation of the Fermi
polaron \cite{Schirotzek2009,Koschorreck2012,Kohstall2012,Wenz2013},
the emergenge of pairing
\cite{BCS1957,Legget2006,Regal2003,Bartenstein2004,Zwierlein2005,Giorgini2008,Zuern2013}. This has fueled theoretical proposals that
are specific to few-atom traps, focusing on themes as diverse
as Stoner ferromagnetism \cite{Liu2010a,Bugnion2013b,Lindgren2013,Cui2013,Sowinski2013,Gharashi2013},
exchange mechanisms \cite{Bugnion2013,Volosniev2013},
Wigner localization \cite{Cremon2010,Wang2012b},
quasiparticle- \cite{Rontani2012} and pair-tunneling \cite{Rontani2013b},
the Fulde-Ferrell-Larkin-Ovchinnikov
state \cite{Bugnion2013c}, pairing \cite{Rontani2009c}, mesoscopic phase
separations for bosons \cite{Garcia-March2013}.
On one hand, the study of few interacting atoms, 
which are the bulding blocks of 
many-body states, gives an insight into
the essential physical features 
of more complex quantum systems. 
On the other hand, the few-body problem may be analyzed using
fully understood theoretical models \cite{Blume2012,Zinner2013} 
that provide a benchmark for approximate theories fit to larger 
numbers of particles. 

In this paper we report accurate calculations
of the energy spectra of few atoms interacting through contact forces
in a one-dimensional harmonic trap. 
Since an analytic 
solution (recalled in section~\ref{formulation})
is available for two particles \cite{Busch1998}, we consider
three atoms, which are either spinful fermions or spinless bosons.
Our approach is based on a 
well controlled variational method (VM)---inspired by previous work in 
two \cite{Liu2010b} and
three \cite{Liu2010a} dimensions---that is physically transparent,
since the three-atom basis set is made of the
exact two-body wave functions \cite{Busch1998} 
plus one-body spectator orbitals. 
We then use the VM spectrum as a benchmark for state-of-the-art  
exact diagonalization [also known as full configuration
interaction (CI)], which is the technique of choice 
for accurate calculations of few-body systems. The CI basis set is simpler
than the VM one, as it is made 
of the Slater determinants
obtained by filling with three fermions a truncated set of trap orbitals,
consistently with Pauli exclusion principle. The comparison
between VM and CI data
points to the slow convergence of CI in the regime of 
strong attractive interactions,
highlighting the challenging nature of theories of pairing in few Fermi-atom
systems. 

The CI technique uniquely accesses energies and wave functions
of both ground and excited states, at the price of being limited to few
atoms due to the exponential increase of the
Hilbert space size with the number of particles \cite{Reimann2002,Rontani2006}.
Together with quantum Monte Carlo \cite{Ceperley1995,Astrakharchik2004,Giorgini2008,Casula2008,Astrakharchik2013}, 
CI provides the ground-state 
energy with---in principle--arbitrary accuracy, hence
it has been widely applied to one-dimensional
systems of finite size 
\cite{Bryant1987,Hausler1993,Reimann2002,Szafran2004,Fiete2005,Gindikin2007,Deuretzbacher2007,Yin2008,Secchi2009,Secchi2010,Kristinsdottir2011,Secchi2012,Brouzos2012,Wang2012b,Pecker2013,Secchi2013,Cui2013,Bugnion2013,Bugnion2013b,Bugnion2013c,Sowinski2013}.
However, for strong attractive interactions,
the two-body wave function tends to collapse 
in space \cite{Busch1998}, hence the convergence of the calculation
should be carefully assessed.
An accurate calculation of the spectrum
of three interacting fermions, based on Green function's formalism,
was provided by Blume and coworkers \cite{Gharashi2012,Gharashi2013}. 
These authors investigated the dimensional crossover from 
a three-dimensional to a quasi one-dimensional trap but reported 
partial data for stricly one-dimensional fermions
and no data for bosons. 
Other approaches for three particles in a trap
in one dimension include group-theoretical  
\cite{Harshman2012} and geometrical analyses \cite{Wilson2013},
multiconfigurational time-dependent Hartree method 
\cite{Zollner2006,Ernst2011,Brouzos2012,Brouzos2013},
ansatz correlated wave functions \cite{Brouzos2012,Brouzos2013},
effective-interaction approaches \cite{Ernst2011,Lindgren2013},
density functional theory \cite{Gao2006},
as well as exact results in the limit of infinite 
repulsion \cite{Guan2009,Girardeau2010b,Volosniev2013}.

This article is organized as follows: 
in section~\ref{formulation} we work out the analytical solution for two atoms 
based on the Bethe-Peierls boundary condition,
as a preliminary to the three-body problem that is introduced in 
section~\ref{s:3B}. 
In sections~\ref{3-fermions-problem} and \ref{3-bosons-problem} 
we derive the energy spectra for fermions and bosons, respectively, 
focusing on the limit of strong repulsion in
section~\ref{fermionization}.
In section~\ref{results_analysis} we compare three-fermion energies with
those obtained from CI calculations employing different single-particle
basis sets. After the conclusion (section~\ref{s:conclusion}), 
\ref{two_body_solution} presents an alternative derivation
of the two-body solution and \ref{A_{pq}} explains the
calculation of the matrix elements that occur in the equations
for the three-body problem.

\section{Two-body problem}\label{formulation}

As a preliminary step, it is convenient to recall 
the exact solution of the two-body problem 
with contact interaction in a one-dimensional trap. 
Whereas in \ref{two_body_solution} we work out this solution
following the original calculation by Busch
and coworkers \cite{Busch1998}, 
in this section we present an alternative approach based on the application
of the Bethe-Peierls boundary condition. We will extend this method 
to three atoms in sections~\ref{3-fermions-problem}
and \ref{3-bosons-problem}.

In a trap with tight transverse
confinement, the low-energy dynamics is effectively one-dimensional.
The hamiltonian $H_{2p}$ for two particles has the following form:
\begin{equation} \label{ham-normal}
H_{2p} = \frac{\bar{p}^2_1 + \bar{p}^2_2}{2m} + 
\frac{1}{2}m\omega^2(\bar{x}^2_1+\bar{x}^2_2)
+\bar{g}\delta(\bar{x}_1-\bar{x}_2),
\end{equation}
where $m$ is the atom mass, $\omega$ is the frequency of the harmonic 
oscillator, $\bar{g}$ is the interaction strength of the contact interaction,
$\bar{p}=i\hbar\partial /\partial \bar{x}$. 
Throughout the paper we place bars over certain quantities having physical
dimensions to discriminate them from their 
dimensionless counterparts. Physically, the contact interaction
is a pseudopotential which mimics 
Van der Waals inter-atomic interaction at low energies \cite{Pethick2001}.
The handle to tune $\bar{g}$ is
the relation \cite{Olshanii1998}
\begin{equation}
\bar{g} = \frac{ 2\hbar^2 a_{{\rm 3D}} }{\mu \ell^2_{\perp} }
\frac{1}{1 - C a_{{\rm 3D}} / \ell_{\perp} },
\label{eq:CIR}
\end{equation}
where $\mu = m/2$ is the reduced mass, $a_{{\rm 3D}}$ is the
three-dimensional scattering length, $\ell_{\perp}$ is the
harmonic-oscillator length in the transverse direction,
$C=1.4603\ldots=\zeta(1/2)$ with $\zeta$ being Riemann's zeta function.
A change in the magnetic offset field through a Feshbach resonance modifies
$a_{\rm 3D}$, and hence $\bar{g}$. It is clear from \eref{eq:CIR}
that the vanishing of the denominator---a resonance due to the
tranverse confinement---allows to reach the ``unitarity'' limit,
$\left|\bar{g}\right|\rightarrow \infty$, going from $\bar{g}=-\infty$
to $\bar{g}=\infty$ or vice versa by a small field variation.

Because the interaction in Hamiltonian \eref{ham-normal}
is short-ranged, it affects only two fermions having opposite spins,
hence the solution is the same as the one for spinless bosons.
It is useful to decouple
center-of-mass and relative motions by introducing 
the coordinates $\bar{X}=(\bar{x}_1+\bar{x}_2)/2$ 
and $\bar{x}=\bar{x}_1-\bar{x}_2$, hence the total Hamiltonian
is the sum of two terms,
\begin{equation} \label{ham-reduced}
H_{2p} = H_{\bar{X}} + H_{\bar{x}},
\end{equation}
with
\begin{equation}
H_{\bar{x}} = \frac{\bar{p}^2}{2\mu} + 
\frac{1}{2}\mu\omega^2 \bar{x}^2 +\bar{g}\delta(\bar{x}),
\end{equation}
and
\begin{equation}
H_{\bar{X}} = \frac{\bar{P}^2}{2M} + \frac{1}{2}M\omega^2 \bar{X}^2,
\label{Hcm}
\end{equation}
where the center-of-mass term \eref{Hcm} is a non-interacting 
harmonic oscillator with doubled mass $M = 2m$ and energy 
$E_{\bar{X}}=\hbar\omega(n+1/2)$, with $n=0,1,2,\dots$
Furthermore, by using $\hbar\omega$ as energy unit and 
$\ell = (\hbar/ \mu \omega)^{1/2}$ as length unit, we introduce 
in the relative-motion frame
the dimensionless variables $\bar{E}_{\bar{x}}=(\hbar\omega) E_x$, 
$\bar{g}=(\hbar\omega \ell)g$, $\bar{x}=\ell x$, 
$\bar{p}=\left(\hbar/\ell\right)p$, with $p=i\partial/\partial x$. 
Therefore, the total wave function is the product of the non-interacting 
center-of-mass oscillator  
times the interacting wave function $\psi(x)$ that obeys the eigenvalue
equation
\begin{equation} \label{eigenproblem-dimensionless}
 \left[\frac{p^2}{2} + \frac{x^2}{2} + g\delta(x)\right]\psi(x) = E_x \psi(x).
\end{equation}

There are two ways to solve the eigenvalue problem  
\eref{eigenproblem-dimensionless}.
One way is to expand the wave function $\psi(x)$ over the eigenstates of
the non-interacting problem, i.e. the states of the
harmonic oscillator
\begin{equation} \label{harmonic_oscillator_solution}
 R_n(x) = \frac{1}{\sqrt{2^n n!}}\left( 
\frac{1}{\pi}\right)^{\frac{1}{4}}\rme^{-\frac{x^2}{2}}H_n\!\left(x\right),
\end{equation}
with $H_n(x)$ being the Hermite polynomial of order $n$. 
This approach follows the
original derivation by Busch and coworkers \cite{Busch1998} and is
detailed in \ref{two_body_solution}.
In this section we take a different path and match the generic 
solutions of \eref{eigenproblem-dimensionless}
in the two half-spaces $x<0$ and $x>0$ by means of the 
(Bethe-Peierls) contact condition \cite{Bethe1935},
\begin{equation} \label{1DBethe-PeierlsCondition}
 \frac{\partial \psi(0^+) }{\partial x} - 
\frac{ \partial \psi(0^-) }{\partial x} = 2g \psi(0).
\end{equation}

We write the generic solution of Eq.~\eref{eigenproblem-dimensionless}
with $x\neq 0$ as  
\begin{equation} \label{ansatz}
\psi(x) =  e^{-\frac{x^2}{2}}f(x),
\end{equation}
with $f(x)$ to be determined.
Setting the relative-motion energy to $E_x = (2\nu +1/2)$ the
eigenvalue problem \eref{eigenproblem-dimensionless} is
reduced to  
\begin{equation}
\frac{ \partial^2 f(x)}{\partial x^2} -2x
\frac{\partial f(x)}{\partial x} + 4 \nu f(x) = 0.
\end{equation}
With the transformation $z = x^2$ we obtain
\begin{equation} \label{kummer_equation}
z\frac{\partial^2 f(z)}{\partial z^2} +
\left( \frac{1}{2}-z\right)
\frac{\partial f(z)}{\partial z} + \nu f(z) = 0.
\end{equation}
Expression (\ref{kummer_equation}) is Kummer's equation 
with parameters $a=-\nu$ and $b=1/2$, 
the solutions being Kummer's functions 
$M(-\nu,1/2,x^2)$ and $U(-\nu,1/2,x^2)$ \cite{Abramowitz1964}. 
The contact condition (\ref{1DBethe-PeierlsCondition}) 
imposes that the solution has a singular derivative at the origin. 
Since $M(-\nu,1/2,x^2)$ is proportional to the Hermite polynomial
and hence has continuous derivatives at $x=0$ whereas those
of $U(-\nu,1/2,x^2)$ are singular, 
the generic solution in the whole space is
\begin{equation} \label{ansatz_2}
\psi(x) = \rme^{-\frac{x^2}{2}}U\!\!\left( -\nu,\frac{1}{2}, x^2 \right),
\end{equation}
that is the same as Eq.~(\ref{2_body_solution_Busch}) except for 
a normalization constant.

To find the eigenvalues, we recall a few properties of Kummer's function 
$U$ in the limit $x\rightarrow 0$ \cite{Abramowitz1964}:
\numparts
\begin{equation}\label{limit_behaviour}
U\!\!\left(-\nu,\frac{1}{2},x^2\right) 
= \frac{\Gamma (1/2)}{\Gamma(-\nu+1/2)} + O(|x|),
\end{equation}
\begin{equation}\label{limit_behaviour_b} 
U\!\!\left(-\nu+1,\frac{3}{2},x^2\right) = 
\frac{1}{|x|}\frac{\Gamma (1/2)}{\Gamma(-\nu+1)} + O(1),
\end{equation}
\begin{equation}\label{limit_behaviour_c} 
\frac{ \partial U\!\!\left(a,b,x\right)}{ \partial x}  
= -a\, U\!\!\left(a+1,b+1,x\right).
\end{equation}
\endnumparts
Using (\ref{limit_behaviour_c}) we obtain the 
wave function derivative:
\begin{eqnarray}
 \frac{\partial}{\partial x}\psi(x) &=& 2x\nu 
U\left(-\nu+1,\frac{3}{2},x^2\right)\rme^{-\frac{x^2}{2}}+ \nonumber \\
&-&  x\psi(x).
\end{eqnarray}
Applying (\ref{limit_behaviour_b}) for vanishing $x$ we get
\begin{equation} \label{non_continuos_derivative}
 \frac{\partial}{\partial x}\psi(x)_{x\rightarrow 0^{\pm}} 
= -2\frac{\sqrt{\pi}}{\Gamma(-\nu)} {\rm sign}(x),
\end{equation}
and using the condition (\ref{1DBethe-PeierlsCondition}) 
we obtain the eigenvalue equation,
\begin{equation} \label{2-body-spectrum}
\frac{\Gamma(-\nu)}{\Gamma\left(-\nu+1/2\right)} = -\frac{2}{g},
\end{equation}
linking the interaction strenght $g$ to the energy $E_x$ since 
$E_x = 2\nu +1/2$.

\begin{figure}[htbp]
\centering
\includegraphics[width=100mm, angle=0]{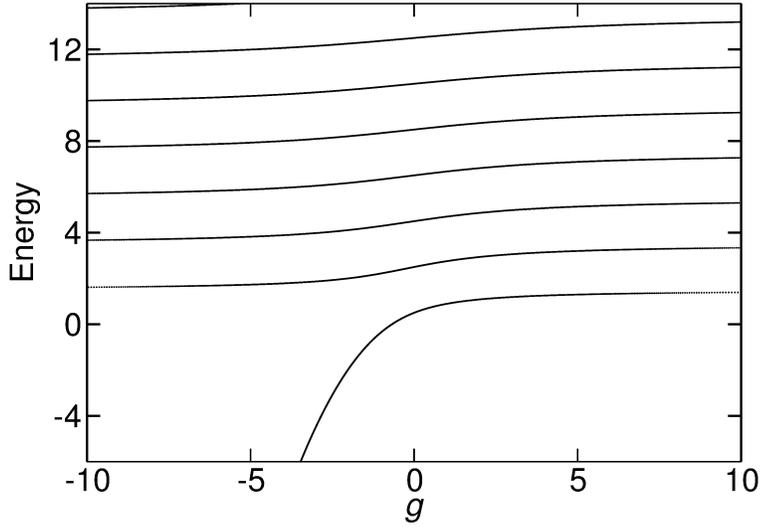}
\caption{Relative-motion energy of two atoms, $E_x$, 
vs interaction strength, $g$.
The energy is in units of $\hbar\omega$ and $g$ in units of $\hbar\omega\ell$.
}
\label{spectrum_two_fermions}
\end{figure}
The resulting energy spectrum, which is plotted 
in figure \ref{spectrum_two_fermions},
exhibits peculiar features.
First, the lowest energy branch drops to $-\infty$ for strong attractive
interactions, the wave function collapsing in space.
Second, the unitarity limit $\left|g\right|\rightarrow \infty$
shows the fermionization of the interacting energy spectrum, which
tends to the noninteracting values that are
peculiar of the spin-polarized system, 
$E_x = 3/2, 7/2, 11/2, \ldots$, whereas the missing values
$E_x = 5/2, 9/2, 13/2, \ldots$ are obtained trhough successive
center-of-mass excitations. 
Indeed, the wave function in the relative
frame is even, displaying a bosoniclike behavior, hence for strong
repulsion it is expected to exhibit the same observable properties as 
fully polarized fermions that are noninteracting.
The same behavior is expected in the presence of strong attraction
but only for `super-Tonks-Girardeau'
excited states \cite{Astrakharchik2005,Guan2009,Girardeau2010,Girardeau2010b}, 
as the lowest collapsing state has no fermionic counterpart.
Experimentally, the ground-state energy branch on the attractive side
($g<0$) was probed for fermions in Ref.~\cite{Zuern2013}
whereas that on the positive side
close to $g\rightarrow \infty$ was measured in Ref.~\cite{Zuern2012}.
The latter experiment also accessed the super-Tonks-Girardeau
first excited energy branch close to $g\rightarrow -\infty$.

\section{Three-body problem}\label{s:3B}

In this section we introduce the three-atom problem and set the basis of our 
variational method, which is inspired by the work of Drummond and coworkers
in two- \cite{Liu2010b} and three-dimensional \cite{Liu2010a} traps. 
In sections~\ref{3-fermions-problem} and \ref{3-bosons-problem} 
we specialize the method to fermions and bosons, respectively.

\begin{figure}[htbp]
\centering
\includegraphics[width=80mm, angle=0]{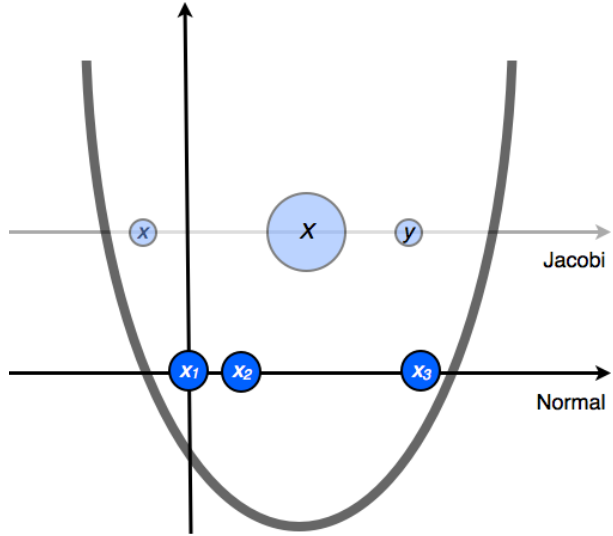}
\caption{Pictorial representation of three atoms in a harmonic trap. 
In the usual laboratory frame
the three atoms (depicted as circles) have like masses.
In the Jacobi reference frame the coordinates represent collective 
motions with equivalent masses, which are depicted as circles whose 
radius sizes are proportional to the masses.}
\label{scheme}
\end{figure}

The Hamiltonian is:
\begin{eqnarray} \label{ham-3_part}
H_{3p} &= & \frac{\bar{p}^2_1 + \bar{p}^2_2+\bar{p}^2_3 }{2m} 
+ \frac{1}{2}m\omega^2\left(\bar{x}^2_1+\bar{x}^2_2+\bar{x}^2_3\right) 
\nonumber \\
 &+& \bar{g}\left[\delta(\bar{x}_1-\bar{x}_2)+
\delta(\bar{x}_1-\bar{x}_3)+\delta(\bar{x}_2-\bar{x}_3)\right].
\end{eqnarray}
After introducing the Jacobi coordinates,
\begin{eqnarray} \label{jacoby_coord}
\bar{X} &=& \frac{\bar{x}_1+\bar{x}_2+\bar{x}_3}{3}, \nonumber \\
\bar{x} &=& \bar{x}_1-\bar{x}_2, \nonumber \\
\bar{y} &=& \frac{2}{\sqrt{3}}\left(\bar{x}_3 
- \frac{\bar{x}_1+\bar{x}_2}{2} \right),
\end{eqnarray}
the operator (\ref{ham-3_part}) is decoupled into three terms:
\begin{equation}\label{ham-3_part_new}
H_{3p} = H_{\bar{X}} + H_{\bar{x}} + H_{\bar{x}\bar{y}},
\end{equation}
with
\begin{eqnarray} 
 H_{\bar{X}} &=& \frac{\bar{P}^2}{2M}+\frac{1}{2}M\omega^2 \bar{X}^2,  
\nonumber \\
H_{\bar{x}} &=&\frac{\bar{p}^2_x}{2\mu}+
\frac{1}{2}\mu \omega^2 \bar{x}^2 + \bar{g}\,\delta(\bar{x}), \nonumber \\
H_{\bar{x}\bar{y}} &=& \frac{\bar{p}^2_y}{2\mu}
+\frac{1}{2}\mu\omega^2 \bar{y}^2 
+\bar{g}\,\delta\!\left(
(\bar{x}-\sqrt{3}\bar{y})/2\right) 
+ \bar{g}\,\delta\!\left((\bar{x}+\sqrt{3}\bar{y})/2\right), 
\end{eqnarray}
and $M = 3m$, $\mu = m/2$.
As for two atoms, the center-of-mass motion is 
a harmonic oscillation decoupled from the relative motion.
Using again as unit length $\ell$ and unit energy $\hbar\omega$,
the eigenvalue problem for the relative motion,
in terms of dimensionless variables, is
\begin{equation}
\left(H_x + H_{xy}\right)\psi_{{\rm rel}}(x,y)=E\,\psi_{{\rm rel}}(x,y).
\label{ham-3_part-dimensionless_0}
\end{equation}
Explicitly, Eq.~\eref{ham-3_part-dimensionless_0} reads as
\begin{equation} \label{ham-3_part_rel}
 \left[\frac{p^2_x}{2}+\frac{x^2}{2} + g\delta(x) + 
\frac{p^2_y}{2}+\frac{y^2}{2} +g\,G(x,y)\right]\psi_{
{\rm rel}} =E\,\psi_{{\rm rel}},
\end{equation}
with the shorthand 
\begin{displaymath}
G(x,y)=\delta\!\left((x-\sqrt{3}y)/2\right) 
+ \delta\!\left((x+\sqrt{3}y)/2\right).
\end{displaymath}
The system is schematically represented in figure \ref{scheme}.

In the absence of the interaction term $G(x,y)$, the solution
of \eref{ham-3_part_rel} is 
the product of the two-atom relative-motion
wave function $\psi(x)$ times the harmonic-oscillator
solution $R_n(y)$. The term $G(x,y)$ couples $x$ and $y$ degrees of 
freedom, hence the generic solution of \eref{ham-3_part_rel} is
\begin{equation} \label{solution3particles}
\Omega(x,y) = \sum_{n=0}^{\infty} a_n \psi_n(x)R_n(y),
\end{equation}
which relies on the completeness of the basis $\{\psi_n(x)R_n(y)\}_n$.
The next two sections are devoted to determine the unknown coefficients $a_n$
in \eref{solution3particles}
after imposing the Bethe-Peierls contact 
condition---the analog of (\ref{1DBethe-PeierlsCondition})---and
the proper symmetry under particle exchange.

In the following we normalize $\psi_n(x)$ as:
\begin{equation}
\psi_n(x) = \Gamma(-\nu_n)\,\rme^{-\frac{x^2}{2}} 
U\!\left(-\nu_n,\frac{1}{2},x^2\right),
\end{equation}
where, for fixed $n$ and $E$, the index $\nu_n$ 
entering the functional form of $\psi_n(x)$ is dictated by
the energy conservation,
\begin{equation} \label{energy_conservation}
E = \left(2\nu_n+\frac{1}{2}\right)+ \left(n+\frac{1}{2}\right) 
= 2\nu_n + n + 1.
\end{equation}
 
\section{Three fermions}\label{3-fermions-problem}

The wave function $\psi_{{\rm rel}}(x,y)\equiv \psi_{3F}(x,y)$ for three
fermionic atoms must fullfill the Pauli principle. Identifing the 
particles with the indices $\{123\}$, we specify the 
spin configuration choosing e.g. 
$\{\uparrow \downarrow \uparrow\}$ (the fully polarized system is
not interacting), hence atoms 1 and 3 are indistinguishable.
Exchanging particles 1 and 3 causes the following coordinate transformation: 
\begin{eqnarray}
x\rightarrow \frac{x}{2} + \frac{\sqrt{3}}{2}y \equiv \xi , \nonumber \\ 
y \rightarrow \frac{\sqrt{3}}{2}x - \frac{y}{2} \equiv \eta.
\label{eq:transf}
\end{eqnarray}
Therefore, the ansatz
\begin{equation} \label{wavefunction3particles}
\psi_{3F}(x,y) = \Omega(x,y) - \Omega\left(\xi,\eta\right)
\end{equation}
has the correct symmetry, since it changes sign under 
the $1 \leftrightarrow 3$ swap. This is immediate
by writing $\psi_{3F}(x,y)$ as
\begin{equation} \label{wavefunction3particles2}
\psi_{3F}(x,y) = (1-{\bm{P}}_{13})\Omega(x,y),
\end{equation}
where ${\bm{P}}_{13}$ is the exchange operator.

The wave function $\psi_{3F}(x,y)$
must also have a definite parity under the inversion operator $\bm{R}$
that changes the sign of all coordinates, 
$\bm{R}\{x_1,x_2,x_3\}=\{-x_1,-x_2,-x_3\}$. 
This condition is easily realized noticing that
the transformation \eref{eq:transf}
is linear and homogeneous, hence
$\bm{R}\{x,y\}=\{-x,-y\}$, and that $\psi_n(x)$ is even in $x$.
Therefore, the inversion leads to
\begin{eqnarray}
&&\bm{R}\,\psi_{3F}(x,y) = \Omega(-x,-y) - \Omega(-\xi,-\eta)  \nonumber \\
                 &&= \Omega(x,-y) - \Omega(\xi,-\eta)  \nonumber \\
                 &&= \sum_{n=0}^{\infty} (-1)^n a_n \left[ \psi_n(x)
                R_n(y)- \psi_n(\xi)R_n(\eta) \right],
\label{eq:inversion}
\end{eqnarray}
showing that the parity of $\psi_{3F}(x,y)$ depens on that of the
harmonic-oscillator states $R_n$ by
choosing only terms with even (odd) indices in the sum \eref{eq:inversion}.

To find the expansion coefficients $a_n$ 
we impose on $\psi_{3F}(x,y)$ the Bethe-Peierls contact condition 
for two fermions of opposite spin approaching each other.
For fermions 1 and 2 one has
\begin{equation} \label{1DBethe-PeierlsCondition_3p}
 \left[\frac{\partial \psi_{3F}(x,y)}{\partial x}\right]_{x\rightarrow 0^+}- 
\left[\frac{\partial \psi_{3F}(x,y)}{\partial x}\right]_{x\rightarrow 0^-} 
 = 2g\, \psi_{3F}(0,y).
\end{equation}
For fermions 2 and 3 one has 
\begin{eqnarray*}
x_2 & \rightarrow & x_3, \nonumber \\
x & \rightarrow & x_1 - x_3, \nonumber \\
y & \rightarrow & \frac{2}{\sqrt{3}}\left[x_3 - \frac{x_1+x_3}{2}\right] 
= \frac{x_3-x_1}{\sqrt{3}} = -\frac{x}{\sqrt{3}},
\end{eqnarray*}
hence
\begin{eqnarray*}
\xi &\rightarrow & \frac{x}{2} - 
\frac{\sqrt{3}}{2}\frac{x}{\sqrt{3}} = 0 \\ \nonumber
\eta &\rightarrow & \frac{\sqrt{3}}{2}x + \frac{x}{2\sqrt{3}} 
= \frac{2}{\sqrt{3}}x .
\end{eqnarray*}
Therefore, the second contact condition is
\begin{equation} \label{1DBethe-PeierlsCondition3}
\left[\frac{\partial \psi_{3F}}{\partial \xi} \right]_{\xi \rightarrow 0^+}-\left[\frac{\partial \psi_{3F}}{\partial \xi} \right]_{\xi \rightarrow 0^-} 
= 2g \left[\psi_{3F} \right]_{\xi=0}.
\end{equation}
It turns out that \eref{1DBethe-PeierlsCondition3} is
automatically satisfied once (\ref{1DBethe-PeierlsCondition_3p})
is enforced due to the exchange symmetry of $\psi_{3F}$.

To apply the condition \eref{1DBethe-PeierlsCondition_3p}
we note that the 
derivative of Eq.~(\ref{wavefunction3particles}) is
\begin{equation} \label{derivative_3F}
 \frac{\partial}{\partial x} \psi_{3F}\left( x, y\right) 
=  \frac{\partial}{\partial x} \Omega\left( x, y\right) 
-\frac{1}{2}\left[ \frac{\partial}{\partial \xi} 
\Omega\left( \xi, \eta\right) + 
\sqrt{3}\frac{\partial}{\partial \eta} \Omega\left( \xi, \eta\right)  \right], 
\end{equation}
where the last two terms on the righ-hand side of the above equation have 
continuos derivative
at $x=0$ and hence do not contribute to the contact constraint. 
We only need the following derivative:
\begin{equation} \label{derivative_3fermions}
 \frac{\partial}{\partial x} \Omega\left( x, y\right) =  
\frac{\partial}{\partial x} \sum_n a_n \psi_n(x)R_n(y)  
= \sum_n a_n \left[ \frac{\partial}{\partial x}\psi_n(x) \right]R_n(y).  
\end{equation}
Using Eqs.~\eref{limit_behaviour_c} and \eref{limit_behaviour_b}
for vanishing $x$ as in section~\ref{formulation} we obtain
\begin{equation} \label{non_continuos_derivative_3p}
 \frac{\partial}{\partial x}\psi_n(x)_{x\rightarrow 0^{\pm}} 
= -2\sqrt{\pi} \,{\rm sign}(x).
\end{equation}
Noting that
\begin{equation} 
\psi_{3F}(x=0) = \Omega(0,y) - 
\Omega\left(\frac{\sqrt{3}}{2}y,-\frac{y}{2}\right),
\end{equation}
expanding it close to $x=0$ using (\ref{limit_behaviour}),
and combining it with (\ref{non_continuos_derivative_3p}), 
the Bethe-Peierls condition \eref{1DBethe-PeierlsCondition_3p} becomes
\begin{eqnarray}\label{1DBethe-PeierlsCondition4}
2g \sum_n a_n \Bigg[\sqrt{\pi}
\frac{\Gamma(-\nu_n)}{\Gamma(-\nu_n+1/2)}R_n\!\left(y\right) 
- \psi_n\!\left(\frac{\sqrt{3}}{2}y\right)R_{n}\!\left(-\frac{y}{2}\right) 
\Bigg] \nonumber \\ =  - 4\sqrt{\pi}\sum_n a_n R_n(y),
\end{eqnarray}
with the indices appearing in \eref{1DBethe-PeierlsCondition4}
being either even or odd depending on the parity of $\psi_{3F}$.
Exploiting 
the orthonormality of the orbitals $R_n(y)$, the
eigenvalue problem \eref{1DBethe-PeierlsCondition4}
may be written as a set of linear equations,
\begin{equation}\label{generalized_eigenvalue}
\sum_n F_{mn}\, a_n = -\frac{2\sqrt{\pi}}{g} a_m,
\end{equation}
with the matrix $F_{mn}$ being defined as
\begin{equation} \label{matrix_A-B}
F_{mn}= \sqrt{\pi}\frac{\Gamma(-\nu_n)}{\Gamma(-\nu_n+1/2)}\delta_{nm} 
- A_{mn},
\end{equation}
with
\begin{equation}
A_{mn}= 
\int_{-\infty}^{\infty} \!\!\!
\rmd y\, R_m\!\left(y\right)\psi_n\!\left(\frac{\sqrt{3}}{2}y\right)
R_{n}\!\left(-\frac{y}{2}\right).
\label{eq:A_nm}
\end{equation}
Note that, when putting $A_{mn}=0$, one recovers the two-body
eigenvalue equation \eref{2-body-spectrum}, as $A_{mn}$ 
is the matrix element of the interaction between the pair 
of atoms in the state $\nu_n$ and the ``spectator'' atom in level $m$.

The eigenvalue $E$ is hidden in the linear system 
\eref{generalized_eigenvalue} through the index $\nu_n
= (E-n-1)/2$ entering $F_{mn}$. 
Practically, to solve the problem we adopt the following procedure:
We first choose a cutoff for the size of the linear system
\eref{generalized_eigenvalue}, with $n = 0,2,4,\ldots,n_{{\rm max}}$
for even parity and $n = 1,3,5,\ldots,n_{{\rm max}}+1$ for odd 
parity, respectively.
We then fix $E$ and hence $\nu_n$, evaluate the matrix $A_{mn}$ as explained in 
\ref{A_{pq}},
solve numerically the
eigenevalue problem \eref{generalized_eigenvalue} to 
obtain the allowed values of $g$, and iterate the procedure for a 
different value of $E$. 
We eventually invert the relation $g(E)$ and find the energy
branches as a function of the interaction strength $g$, 
after removing trivial results corresponding to non-interacting states.
The latter correspond to fully spin-polarized states after a rotation 
in the spin space.

\begin{figure}[htbp]
\centering
\includegraphics[width=100mm, angle=0]{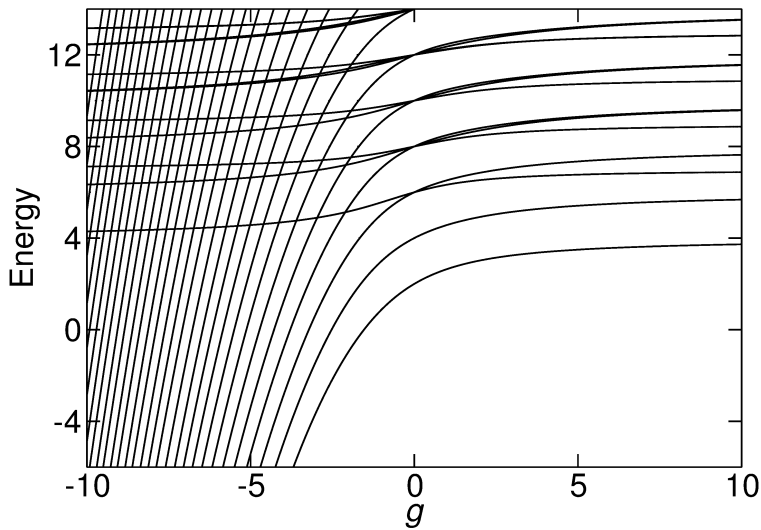}
\caption{Energy of three fermions vs interaction strength $g$ in the relative
frame for states having odd parity. 
The energy is in units of $\hbar\omega$ and $g$ in units of $\hbar\omega\ell$.
}
\label{fermions_odd}
\end{figure}

\begin{figure}[htbp]
\centering
\includegraphics[width=100mm, angle=0]{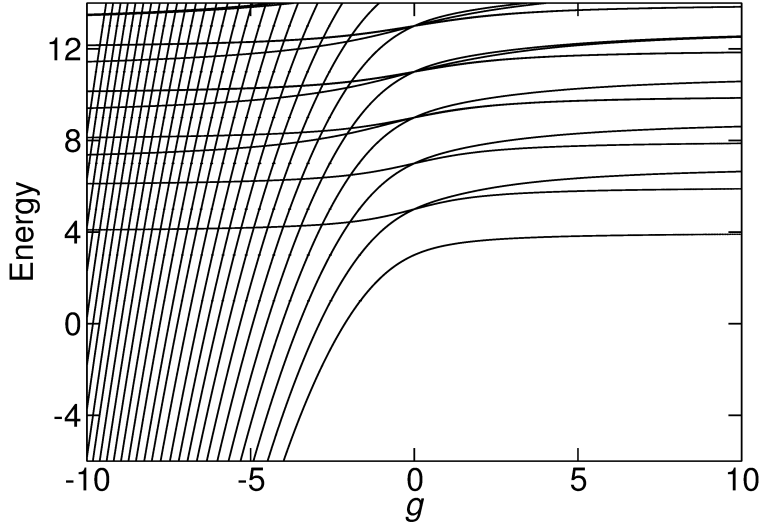}
\caption{Energy of three fermions vs interaction strength $g$ in the relative
frame for states having even parity.
The energy is in units of $\hbar\omega$ and $g$ in units of $\hbar\omega\ell$. 
}
\label{fermions_even}
\end{figure}

Figures \ref{fermions_odd} and \ref{fermions_even}
show the energy spectra for odd and even parity, respectively,
obtained with $n_{{\rm max}}=60$, which allows
the convergence of the sixth digit of the ground-state energy.
With respect to the two-atom spectrum of figure \ref{spectrum_two_fermions},
the plots exhibit new qualitative features. 
First, the orbital states may be either odd (figure \ref{fermions_odd}) or
even (figure \ref{fermions_even}), the former including the
absolute ground state. Second, there are two distinct sets of branches 
in each plot, whose
behavior qualitatively differs as $g\rightarrow - \infty$: 
(i) many branches drop towards $-\infty$ (ii) others tend 
asymptotically to integer values. 
The former may be understood as ground-state replicas 
that are made of two atoms in a strongly-bound state and
the third spectator atom occupying consecutive quasiparticle 
levels of increasing energy. The latter tend to the `fermionized' values
of the fully spin-polarized state
in the unitarity limit \cite{Guan2009,Girardeau2010b,Gharashi2013},
as we will further discuss in section~\ref{fermionization}.
Note that many of the curves shown in figures \ref{fermions_odd} 
and \ref{fermions_even} actually anticross when examined on a finer
energy scale.

\section{Three bosons} \label{3-bosons-problem}

The wave function for three spinless bosons, 
$\psi_{{\rm rel}}(x,y)\equiv \psi_{3B}(x,y)$, must be symmetric under
any particle permutation. The proper ansatz wave function is 
\begin{equation} \label{3BosonsWF}
 \psi_{3B}(x,y) = 
\left( 1+\bm{P}_{13}+\bm{P}_{23} \right)\Omega(x,y),
\end{equation}
as $\Omega(x,y)$ is invariant under the permutation $1\leftrightarrow 2$.
The exchange operator $\bm{P}_{13}$ is the same as in
the ansatz \eref{wavefunction3particles2} for fermions,
whereas $\bm{P}_{23}$ swaps atoms 2 and 3. 
The latter leads to the following coordinate transformation:
\begin{eqnarray}
x\rightarrow \frac{x}{2} - \frac{\sqrt{3}}{2}y \equiv \gamma , \\ \nonumber
y \rightarrow -\frac{\sqrt{3}}{2}x - \frac{y}{2} \equiv \delta.
\end{eqnarray}
Therefore, Eq.~(\ref{3BosonsWF}) may be explicitly written as 
\begin{equation} \label{3BosonsWF1}
 \psi_{3B}(x,y)=\Omega(x,y)+\Omega(\xi,\eta)+\Omega(\gamma,\delta).
\end{equation}

As for fermions, it suffices to impose the contact condition 
(\ref{1DBethe-PeierlsCondition_3p}) for particles 1 and 2, 
\begin{equation} \label{Bethe_bosons}
 \left[\frac{\partial \psi_{3B}(x,y)}{\partial x}\right]_{x\rightarrow 0^+}- 
\left[\frac{\partial \psi_{3B}(x,y)}{\partial x}\right]_{x\rightarrow 0^-} 
 = 2g\, \psi_{3B}(0,y),
\end{equation}
since the conditions for the other pairs are 
automatically satisfied through the exchange symmetry of 
\eref{3BosonsWF}.
The required derivative is:
\begin{eqnarray} \label{3BosonsWF_Derivative}
 \frac{\partial \psi_{3B}(x,y)}{\partial x} &=&  
\frac{\partial \Omega(x,y)}{\partial x}  
+\frac{1}{2}\left[ \frac{\partial \Omega(\xi,\eta)}{\partial \xi} 
+ \sqrt{3}  \frac{\partial \Omega(\xi,\eta)}{\partial \eta}\right]  
\nonumber \\
&+&\frac{1}{2}\left[\frac{\partial \Omega(\gamma,\delta)}{\partial \gamma} 
- \sqrt{3}  \frac{\partial \Omega(\gamma,\delta)}{\partial \delta}\right]. 
\end{eqnarray}
Since the only term that has a singular derivative in $x=0$ is 
$\Omega(x,y)$, one proceeds as in the fermionic case obtaining
\begin{eqnarray} \label{Bethe-Peierls-bosons}
2g\sum_n a_n \Bigg[ \sqrt{\pi} 
\frac{\Gamma(-\nu_n)}{\Gamma(-\nu_n+1/2)}R_n(y)  
+2 \,\, \psi_n\!\left( \frac{\sqrt{3}}{2}y\right)
R_n\!\left(-\frac{y}{2}\right)\Bigg]  \nonumber \\ = 
 -4\sqrt{\pi}\sum_n a_n R_n(y)  .
\end{eqnarray}
Exploiting as before
the orthonormality of the orbitals $R_n(y)$, the
eigenvalue problem \eref{Bethe-Peierls-bosons}
is written as a set of linear equations,
\begin{equation}\label{generalized_eigenvalue_bosons}
\sum_n B_{mn}\, a_n = -\frac{2\sqrt{\pi}}{g} a_m,
\end{equation}
with the matrix $B_{mn}$ being defined as
\begin{equation} \label{matrix_B}
B_{mn}= \sqrt{\pi}\frac{\Gamma(-\nu_n)}{\Gamma(-\nu_n+1/2)}\delta_{nm} 
+ 2A_{mn},
\end{equation}
where $A_{mn}$ is given in \eref{eq:A_nm}
and again the indices $m$ and $n$ assume either even or odd 
values depending on the parity of the bosonic state.
The method to solve the eigenvalue problem 
\eref{generalized_eigenvalue_bosons}
parallels that of section~\ref{3-fermions-problem}.

\begin{figure}[htbp]
\centering
\includegraphics[width=100mm, angle=0]{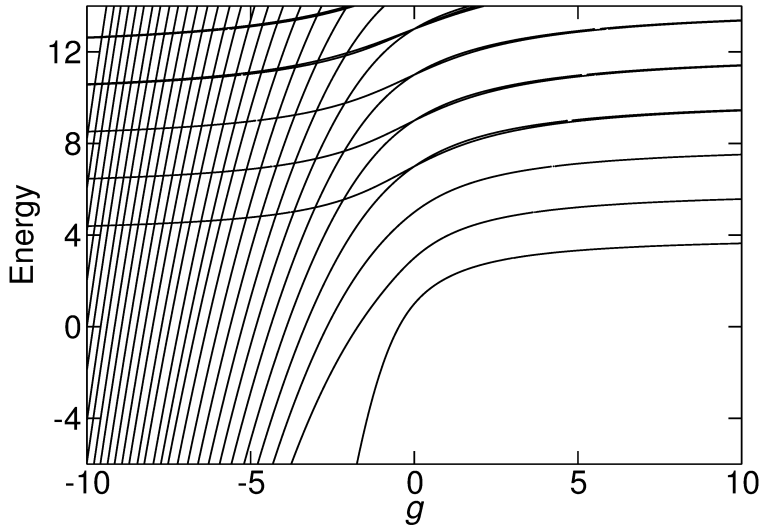}
\caption{Energy of three bosons vs interaction strength $g$ in the relative
frame for states having even parity. 
The energy is in units of $\hbar\omega$ and $g$ in units of $\hbar\omega\ell$.
}
\label{bosons_even}
\end{figure}

\begin{figure}[htbp]
\centering
\includegraphics[width=100mm, angle=0]{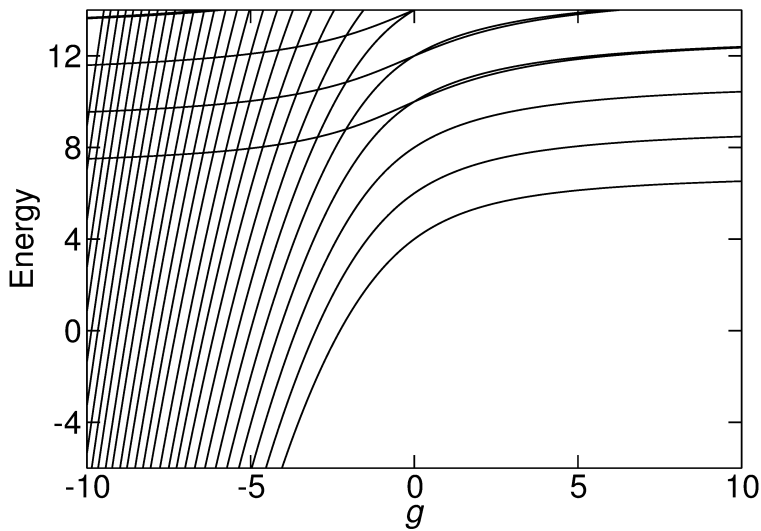}
\caption{Energy of three bosons vs interaction strength $g$ in the relative
frame for states having odd parity. 
The energy is in units of $\hbar\omega$ and $g$ in units of $\hbar\omega\ell$.
}
\label{bosons_odd}
\end{figure}

Figures \ref{bosons_even} and \ref{bosons_odd} show the
energy spectra of three bosons 
obtained with $n_{{\rm max}}=60$, whose wave functions have 
respectively even and odd parities.
The plots are qualitatively similar to those for fermions
(figures \ref{fermions_odd} and \ref{fermions_even}),
displaying both dimer-atom energies that drop to $-\infty$
for increasing attractive interaction as well as energy branches
that tend to fermionized values at unitarity (see also 
section~\ref{fermionization}). Contrary to the fermionic case,
the ground state has now even parity, as it is clear for 
noninteracting particles all filling the lowest harmonic-oscillator state.

\section{Strong repulsion} \label{fermionization}

\begin{figure}[htbp]
\centering
\includegraphics[width=100mm, angle=0]{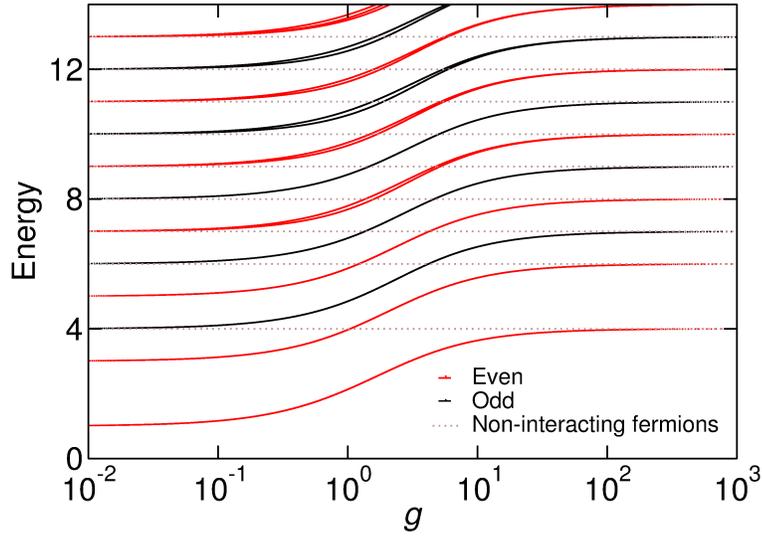}
\caption{Energies of three bosons vs $g$ for large repulsive interaction
in the relative frame. Both even (red [gray] curves)
and odd (black curves) states are considerd.
Note the logarithmic scale of the $g$ axis.  
The dotted horizontal lines point to the energies of three fully
spin-polarized fermions. 
The energy is in units of $\hbar\omega$ and $g$ 
in units of $\hbar\omega\ell$.
}
\label{bosons_fermionize}
\end{figure}

\begin{figure}[htbp]
\centering
\includegraphics[width=100mm, angle=0]{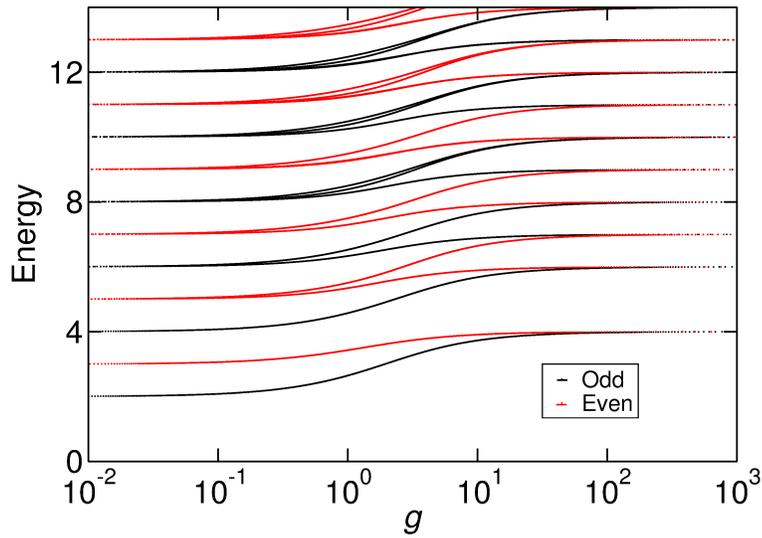}
\caption{Energies of three fermions vs $g$ for large repulsive interaction
in the relative frame. Both even (red [gray] curves) and
odd (black curves) states are considerd.
Note the logarithmic scale of the $g$ axis.
The energy is in units of $\hbar\omega$ and $g$ 
in units of $\hbar\omega\ell$.
}
\label{even_odd}
\end{figure}

In the limit of strong repulsion, $g\rightarrow \infty$, 
bosons are predicted to fermionize, exhibiting the same observable
properties as those of the dual system of fully spin-polarized
fermions \cite{Girardeau1960}. This is evident from 
figure \ref{bosons_fermionize}, which displays the energy levels of
three bosons as $g$ increases up to $10^3$. In this limit, all
branches (continuous curves)
tend to the noninteracting values of the system of three
fermions having parallel spins (horizontal dashed lines), $E=4,6,7,8,\ldots$,
confirming the accuracy of our calculation. Note that the missing value
$E=5$ is obtained when exciting a quantum of the center-of-mass motion.

The behavior of strongly repulsive fermions that are not spin-polarized  
is more involved, due to the emergence of a large degeneracy of levels 
at unitarity \cite{Guan2009,Girardeau2010b,Gharashi2013,Bugnion2013b,Lindgren2013,Sowinski2013,Cui2013,Volosniev2013}. This
may be seen in figure \ref{even_odd}, which shows that 
as $g\rightarrow \infty$ all energy levels
tend to integer values that are twice or more degenerate asymptotically.
Each one of these manifolds at $g=\infty$ includes at least one 
even (red [gray] curves) and one odd (black curves) state. 

To rationalize this trend, consider the
wave function of three fermions in the standard frame, 
$\Psi(x_1,\sigma_1;x_2,\sigma_2;x_3,\sigma_3)$, where $x_i$
is the coordinate of the $i$th atom and $\sigma_i = \pm 1$ 
its spin projection.
The contact condition enforcing Pauli exclusion principle is that
$\Psi\equiv 0$ when $x_i=x_j$ and $\sigma_i=\sigma_j$, for any $i,j$.
In addition, at unitarity one has $\Psi\equiv 0$ also when
$x_i=x_j$ and $\sigma_i=-\sigma_j$ \cite{Guan2009}. These two conditions,
together with the mirror symmetry of the problem, imply that
at unitarity both even and odd states exhibit like nodal surfaces,
owning the same probability density and energy.

\section{Comparison with full configuration 
interaction}\label{results_analysis}

We have tested the results of our variational method (VM) with the data
available in the literature. In particular, the energies of 
three fermions with odd parity agree with the values 
tabulated in Ref.~\cite{Gharashi2013} (Supplemental Material)
to the sixth digit, the dataset including the ground state branch
for $0<g<\infty$ and the super-Tonks-Girardeau branch
from $E=4$ to $E=6$ for $-\infty < g < 0$.
On the other hand, our VM data for the three-fermion ground state 
energy significantly depart from those of Ref.~\cite{Harshman2012}. 
After using the GNU sofware Plot Digitizer and considering
the different unit length definition, we extract from figure 5 of 
Ref.~\cite{Harshman2012} the values 
$E=0.44$, 0.87, 1.25, 1.61, 2, 2.24, 2.48, 2.67, 2.84 for 
increasing values of $g$ going from $g= -2.82843$ to $g=2.82843$ 
in steps of $1/\sqrt{2}$, while the VM energies 
are $E=-3.08$, -1.21, 0.21, 1.26, 2, 2.49, 2.82, 3.05, 3.21 
(cf.~figure \ref{fermions_odd}). The difference is huge especially at
large negative values of $g$. 

\begin {table}
\caption {Size of CI subspace for three fermions vs
single-particle basis set. The data refer to the odd-parity sector
of total spin projection $S_z = \hbar / 2$.
}
\label{hilbert_space_dimensions}
\begin{indented}
\item[]\begin{tabular}{@{} l  l }
\br
    Number of orbitals & CI subspace size \\
\mr
    10 & 450 \\
    25 & 7500 \\
    30 & 13\,050 \\
    35 & 20\,825 \\
    40 & 31\,200 \\
    50 & 61\,250 \\
\br
   \end{tabular}
\end{indented}
\end {table}

Using our VM data as a benchmark, in the following we discuss 
the convergence of the full CI method (also known as exact diagonalization), 
which is widely used for accurate calculations of energies 
and wave functions of few-body systems. The standard CI 
subspace for three fermions is spanned by the Slater determinants obtained 
by filling with three atoms---in all possible ways and consistently
with Pauli exclusion principle---a truncated set of single-particle
harmonic-oscillator orbitals.
Table \ref{hilbert_space_dimensions} reports the size of this CI subspace 
as the number of harmonic-oscillator orbitals increases.  
The full CI method provides a numerically exact solution
in the limit of a complete single-particle basis set, 
the tradeoff for using a truncated basis set being the exponential growth
of the CI subspace, as shown in table \ref{hilbert_space_dimensions}.

\begin{figure}[htbp]
\centering
\includegraphics[width=100mm, angle=0]{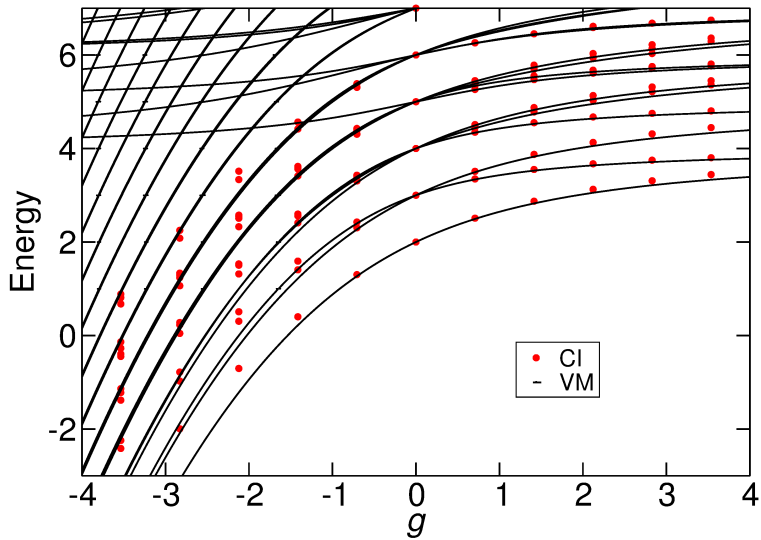}
\caption{Energy spectrum of three fermions vs interaction strength $g$ in
the relative frame including both even and odd states.
The total spin projection is $S_z=\hbar/2$. The data obtained from
the variational method presented in this paper (VM, solid curves)
are compared with those of a full-configuration-interaction calculation
(CI, points). The latter method, also
known as exact diagonalization, employed 35 harmonic-oscillator orbitals. 
The first two excitations of the
center-of-mass motion from the absolute ground state
were included in the plot to facilitate the
comparison between the two techniques.
}
\label{DonRodrigo_comparison}
\end{figure}

Figure \ref{DonRodrigo_comparison} compares the VM energy spectrum for three fermions
(solid curves) with selected CI data obtained from the home-made parallel code 
\emph{DonRodrigo} \cite{Rontani2006} using 35 harmonic-oscillator orbitals (points).
Since \emph{DonRodrigo} uses the standard reference frame its output contains 
both relative-motion and center-of-mass excitations. To facilitate the comparison
between CI and VM data, we have added the first
two center-of-mass excitation quanta to the VM ground-state energy and included both
parities in figure \ref{DonRodrigo_comparison}. 
CI predictions nicely match VM results for repulsive 
interactions, the smaller $g$ the better the agreement. This converging behavior
is generic to long-range repulsive interactions in one \cite{Bryant1987,Hausler1993,Secchi2009,Secchi2010,Wang2012b,Secchi2012,Pecker2013,Secchi2013} and two dimensions 
\cite{Reimann2002,Rontani2006,Climente2007,Kalliakos2008,Singha2010,Gamucci2012}. 
However, the performance of the CI method is poor on the
attractive side, with significant deviations from the VM predictions say for $g < -1.5$. 

\begin{table}
\caption {Ground-state energies of three fermions in the relative frame
for different interaction strengths $g$. 
The reference data obtained from the variational method (VM) presented 
in this work are compared with the results of 
full CI calculations using single-particle basis sets spanned by
$n_{{\rm SP}}$ harmonic-oscillator orbitals. The size of the single-particle
basis set, labeled as `CI $n_{\rm SP}$', is indicated in the first row.
}
\label{data_comparison_with_don_rodrigo}
\begin{indented}
\lineup
\item[]    \begin{tabular}{@{} l l l l l l l }
\br
    $g$       & VM    & CI 50    &   CI 40  & CI 35   & CI 30   & CI 25 \\ 
\mr
    -2.8284 &  -3.0865 &  -2.1476 & -2.0535  & -1.9937 & -1.9216 & -1.8320  \\ 
    -2.1213 &  -1.2165 &  -0.7806 & -0.7340  & -0.7040 & -0.6675 & -0.6217  \\
    -1.4142 &   0.2185 &   0.3723 &  0.3897  &  0.4010 &  0.4149 &  0.4327  \\ 
    -1.0606 &   0.7890 &   0.8637 &  0.8725  &  0.8782 &  0.8853 &  0.8944  \\ 
    -0.7071 &   1.2685 &   1.2976 &  1.3011  &  1.3033 &  1.3062 &  1.3098  \\
    -0.3535 &   1.6695 &   1.6754 &  1.6762  &  1.6767 &  1.6773 &  1.6781  \\ 
     0.3535 &   2.2715 &   2.2755 &  2.2760  &  2.2764 &  2.2769 &  2.2775  \\ 
     0.7071 &   2.4935 &   2.5074 &  2.5093  &  2.5105 &  2.5120 &  2.5141  \\ 
     1.0606 &   2.6755 &   2.7016 &  2.7051  &  2.7074 &  2.7103 &  2.7142  \\ 
     1.4142 &   2.8255 &   2.8639 &  2.8691  &  2.8725 &  2.8769 &  2.8828  \\ 
     2.1213 &   3.0530 &   3.1138 &  3.1220  &  3.1275 &  3.1345 &  3.1439  \\ 
     2.8284 &   3.2140 &   3.2919 &  3.3024  &  3.3095 &  3.3186 &  3.3306  \\ 
\br
   \end{tabular}
\end{indented}
\end {table}

The slow convergence of the CI calculation 
for attractive interactions is detailed
in table \ref{data_comparison_with_don_rodrigo}. The CI single-particle
basis set is spanned by $n_{{\rm SP}}$ harmonic-oscillator orbitals. 
Increasing $n_{{\rm SP}}$ lowers systematically the ground-state energy as one 
moves from the rightmost column of table \ref{data_comparison_with_don_rodrigo}
towards left for a certain value of $g$. However, whereas 
doubling $n_{{\rm SP}}$, going from 25 to 50, allows 
one to match the VM reference
value at least with two-digit precision for all values of $g>0$, 
this accuracy
is obtained only for $g > -0.3535$ on the attractive side.

The reason is that attractive interactions squeeze 
the wave function probability 
weight at lower values of the relative distance $x$ between two atoms.
Therefore, the accuracy of the calculation crucially depends 
on the capability of the
CI wave function to mimic the cusp of the exact wave function at $x=0$, which
occurs as an effect of the Bethe-Peierls contact condition.
Since at the origin the exact
wave function is not analytic, the CI single-particle
basis set, spanned by smooth Hermite polynomials, badly performes and requires
a large cutoff $n_{{\rm SP}}$ to reach reasonable accuracy.
On the contrary, in the VM the wave function is expanded on a basis 
that includes the cusp from the beginning. 

Among the strategies to improve the performance of the CI 
calculation, one possible route is to choose a different
single-particle basis set.
For example, using the eigenstates of the harmonic oscillator
in the relative-motion frame
we found a significant enhancement of convergence for two fermions.
However, this path may be unpractical for more fermions, 
due to the complexity of two-body interaction matrix elements
represented in this basis. Some authors proposed effective-interaction
schemes or similar techniques borrowed by nuclear 
physics \cite{Kvaal2007,Stetcu2007,Alhassid2008,Christensson2009}. 
The interested reader may consult
the recent review \cite{Mitroy2013}. 

\section{Conclusion}\label{s:conclusion}

In this work we have obtained accurate energies for three
particles interacting through contact forces in a one-dimensional
harmonic trap. These results reproduce the exact beahvior
in the limit of infinite interaction and may be used as a benchmark
for other approximate theoretical treatments that are useful for 
larger systems. In particular, we find that the method of exact
diagonalization, also known as full configuration interaction,
converges slowly for strong attractive interactions
between fermions. This points to the challenge of a 
numerical theory of pairing in finite one-dimensional systems that
are currently under experimental investigation.

\ack
We thank Gerhard Z\"urn, Steffi Reimann, and Sven {\AA}berg 
for stimulating discussions.
This work is supported by the EU-FP7 Marie Curie initial training 
network INDEX and by the CINECA-ISCRA grants IscrC\_TUN1DFEW 
and IscrC\_TRAP-DIP.

\appendix

\section{Alternative solution of the two-body problem}\label{two_body_solution}

In this appendix we solve the eigenvalue
problem (\ref{eigenproblem-dimensionless}) in a manner alternative to
that of section~\ref{formulation}. Following the strategy
concisely explained in \cite{Busch1998},
we write the relative-motion wave function 
as a linear combination of the harmonic-oscillator solutions $R_n(x)$
[cf.~\eref{harmonic_oscillator_solution}]:
\begin{equation} \label{solution}
\psi(x) = \sum_{n=0}^{\infty} c_n R_n(x).
\end{equation}
Inserting the representation (\ref{solution}) 
into the Hamiltonian (\ref{eigenproblem-dimensionless}) we obtain
\begin{equation} \label{coefficients}
\sum_{n=0}^{\infty} c_n (\epsilon_n - E_x)R_n(x) = -g \delta(x)\sum_{n=0}^{\infty} c_nR_n(x),
\end{equation}
with $\epsilon_n = n + 1/2$,
and projecting on the eigenstate $R_n(x)$ we get
\begin{equation} \label{coefficients1}
c_n (\epsilon_n - E_x) + 
g R_n(0) \left[\sum_{m=0}^{\infty} c_mR_m(x)\right]_{x\rightarrow 0}=0,
\end{equation}
that can be written as
\begin{equation} \label{coefficients2}
c_n  = A \frac{R_n(0) }{\epsilon_n - E_x},
\end{equation}
with $A$ being an unknown constant.
We need only the even harmonic-oscillator solutions because the 
odd ones have value zero at $x=0$ and correspond to noninteracting particles.
Combining (\ref{coefficients1}) and (\ref{coefficients2}) 
we obtain the identity
\begin{equation} \label{coefficients3}
-\frac{1}{g} = \left[\sum_{n=0}^{\infty} 
\frac{R_n(0)R_n(x)}{E_n-E_x} \right]_{x\rightarrow 0}.
\end{equation}

The harmonic-oscillator orbitals 
may be written in terms of the Laguerre polynomials:
\begin{equation} \label{harmonic-solution-laguerre}
 R_{2n}(x)= \frac{1}{\sqrt{\pi}}\frac{1}{R_{2n}(0)}
\rme^{-\frac{x^2}{2}}L_n^{-\frac{1}{2}}\!\!\left(x^2\right).
\end{equation}
Using this relation (\ref{coefficients3}) takes the form
\begin{eqnarray} \label{coefficients4}
-\frac{1}{g} &=& \left[\frac{1}{\sqrt{\pi}}
\rme^{-\frac{x^2}{2}}\sum_{n=0}^{\infty} 
\frac{L_n^{-\frac{1}{2}}\!\!\left(x^2\right)}{\epsilon_{2n} - E_x} 
\right]_{x\rightarrow 0} = 
\nonumber \\ 
&=& \frac{1}{\sqrt{\pi}}\left[\rme^{-\frac{x^2}{2}}
\sum_{n=0}^{\infty} 
\frac{L_n^{-\frac{1}{2}}\!\!\left(x^2\right)}{2n+1/2 - E_x}
\right]_{x\rightarrow 0} = \nonumber \\
&=& \frac{1}{2\sqrt{\pi}} \left[\rme^{-\frac{x^2}{2}}
\sum_{n=0}^{\infty} \frac{L_n^{-\frac{1}{2}}\!\!\left(x^2\right)}{n-\nu}
\right]_{x\rightarrow 0} ,
\end{eqnarray}
where $\nu = E_x/2 -1/4$ and the dummy index 
runs over all $n$ (even and odd) in the sum.
To calculate the sum 
$\sum_n L_n^{-1/2}(x^2)/(n-\nu)$
we use the two following identities 
(cf.~entry 22.9.15 in \cite{Abramowitz1964}):
\begin{equation} 
\frac{1}{n-\nu} = 
\int_0^{\infty}\!\!\!\frac{\rmd y}{(1+y)^2}\left(\frac{y}{1+y}\right)^{n-\nu-1},
\end{equation}
\begin{equation} 
\sum_{n=0}^{\infty}z^nL_n^{\alpha}(x) = 
(1-z)^{-\alpha-1}\exp\left( \frac{xz}{z-1}\right);
\end{equation}
plus the tabulated integral
(cf.~entry 13.2.5 in \cite{Abramowitz1964}):
\begin{equation}
\int_0^{\infty}\!\!\!\frac{\rmd t}{(1+t)^{a+1-b}}
\,t^{a-1}\rme^{-zt} = \Gamma(a)\,U(a,b,z).
\end{equation}
The final result is:
\begin{equation} \label{sum_laguerre_polynomials}
\sum_{n=0}^{\infty} \frac{L_n^{-\frac{1}{2}}\!\!\left(x^2\right)}{n-\nu}  
= \Gamma(-\nu)\,U\!\!\left(-\nu,\frac{1}{2},x^2\right).
\end{equation}

From \eref{limit_behaviour} the limit of 
Kummer function $U(-\nu,1/2,x^2)$ for $x\rightarrow 0$ is 
\begin{equation}
\lim_{x\rightarrow 0}
U\!\!\left(-\nu,\frac{1}{2},x^2\right) = 
\frac{\Gamma(1/2)}{\Gamma\left(-\nu+1/2\right)},
\end{equation}
hence \eref{coefficients4} becomes
\begin{equation}
-\frac{1}{g} = \frac{1}{2\sqrt{\pi}}\Gamma(-\nu)
\frac{\Gamma(1/2)}{\Gamma\left(-\nu+1/2\right)}.
\end{equation}
Therefore,
we recover the eigenvalue equation \eref{2-body-spectrum},
and plugging the identities 
(\ref{coefficients2}),
(\ref{harmonic-solution-laguerre}) and 
(\ref{sum_laguerre_polynomials}) into the expansion (\ref{solution}) 
we obtain the explicit form of the wave function:
\begin{equation} \label{2_body_solution_Busch}
\psi(x)  = \frac{A}{2\sqrt{\pi}}\,\rme^{-\frac{x^2}{2}}
\,\Gamma(-\nu)\,U\!\!\left(-\nu,\frac{1}{2},x^2\right),
\end{equation}
where the constant $A$ is fixed by the normalization condition.

\section{Explicit form of matrix element $A_{pq}$} \label{A_{pq}}

We rewrite the matrix element 
\eref{eq:A_nm} to evaluate: 
\begin{equation}
A_{pq} = \int_{-\infty}^{+\infty}\!\!\! \rmd y R_p(y)
\,\psi_q\!\!\left( \frac{\sqrt{3}}{2}y\right)R_q\!\!\left(-\frac{y}{2}\right). 
\end{equation}
Writing the harmonic oscillator orbital as
\begin{equation}
R_n(z) = C_n \rme^{-\frac{z^2}{2}}H_n\!\left(z\right),
\end{equation}
with
\begin{equation}
C_n = \frac{1}{\sqrt{2^n n!}}\left( \frac{1}{\pi}\right)^{\frac{1}{4}},
\end{equation}
the integral reads:
\begin{eqnarray}
A_{pq} &=& \int_{-\infty}^{+\infty}\!\!\! \rmd y\, C_p 
\rme^{-\frac{y^2}{2}}\rme^{-\frac{3}{8}y^2}\Gamma(-\nu_q)C_q 
\rme^{-\frac{y^2}{8}}J_{pq}(y)  \nonumber \\
&=&2 C_p\, C_q\, \Gamma(-\nu_q)\int_{0}^{\infty}\!\!\! \rmd
y\, \rme^{-y^2}J_{pq}(y),
\end{eqnarray}
where 
\begin{equation}
J_{pq}(y)=H_p\!\left(y\right)U\!\left(-\nu_q,1/2,3y^2/4\right) 
H_q\!\left(-y/2\right).
\end{equation}
Using the identity
\begin{equation}
\Gamma(a)U\!\left(a,\frac{1}{2},z\right) 
= \sum_{k=0}^{\infty} \frac{L_k^{-\frac{1}{2}}(z)}{k+a} 
 = \sum_{k=0}^{\infty} \frac{(-1)^k U\!\left(-k,1/2,z\right)}{k!(k+a)}, 
\end{equation}
we rewrite the matrix element in the following way:
\begin{equation}
\label{eq:Ipqk}
A_{pq} = 2 C_p C_q \sum_{k=0}^{\infty} 
\frac{(-1)^k}{k!(k-\nu_q)} \int_{0}^{\infty} \!\!\!\rmd y\, I_{pq}^k(y),
\end{equation}
where we have defined the integrand as
\begin{equation}
I_{pq}^k(y) = 
\rme^{-y^2}H_p\!\left(y\right)
U\!\left(-k,\frac{1}{2},\frac{3}{4}y^2\right) 
H_q\!\left(-\frac{y}{2}\right).
\end{equation}

The expression \eref{eq:Ipqk} has the advantage that 
it depends on the energy only through the
parameter $\nu_q$ that enters the denominator of 
the addendum of the sum over $k$.  
Therefore, one computes
and stores the integrals $\int \rmd y I_{pq}^k(y)$ once for all
while repeating the summation over $k$ for different energies, which
makes the computation fast.
In this way we performed calculations truncating the sum \eref{eq:Ipqk}  
up to $k=30$ and obtaining convergence 
on the sixth digit of the ground state energy.

\end{document}